\documentclass[%
 reprint,
 amsmath,amssymb,
 aps,
]{revtex4-2}

\usepackage{graphicx}
\usepackage{dcolumn}
\usepackage{bm}

\begin{document}

\preprint{APS/123-QED}

\title{Tunable Rectangular Resonant Cavities for Axion Haloscopes}
\author{Ben T. McAllister$^{1,2}$}
\email{bmcallister@swin.edu.au}
\author{Aaron P. Quiskamp$^2$}
\author{Michael E. Tobar$^2$}
\affiliation{$^1$ARC Centre of Excellence for Dark Matter Particle Physics, Swinburne University of Technology, John St, Hawthorn VIC 3122, Australia}
\affiliation{$^2$QDM Lab, Department of Physics, University of Western Australia, 35 Stirling Highway, Crawley WA 6009, Australia}

\date{\today}

\begin{abstract}
Axions are a compelling dark matter candidate, and one of the primary techniques employed to search for them is the axion haloscope, in which a resonant cavity is deployed inside a strong magnetic field so that some of the surrounding axions may convert into photons via the inverse Primakoff effect and become trapped inside the resonator. Resonant cavity design is critical to the sensitivity of a haloscope, and several geometries have been utilised and proposed. Here we consider a relatively simple concept - a rectangular resonant cavity with a tunable wall - and compare it to the standard tuning rod-type resonators employed in the field. We find that the rectangular cavities support similar modes to cylindrical tuning rod cavities, and have some advantages in terms of axion sensitivity and practicality, particularly when moving to higher frequencies which are of great and growing interest in the international axion dark matter community.
\end{abstract}

\maketitle

\section{Introduction}

\begin{figure*}[t]
\centering
	\includegraphics[width=0.75\textwidth]{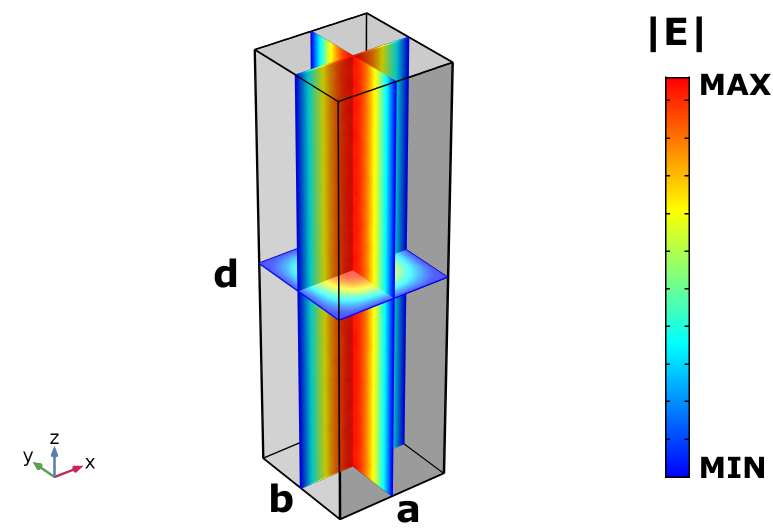}
\caption{A schematic of the exterior of a hollow rectangular resonant cavity with the dimensions $a$, $b$ and $d$ defined along the relevant axes. The electric field magnitude of the TM$_{110}$ mode, the most axion sensitive mode inside the empty rectangular resonator is shown in the rainbow plot. Image adapted from COMSOL Multiphysics output.}
\label{fig:rectanglesketch}
\end{figure*}

The nature of dark matter is one of the biggest mysteries in modern science. There are decades of observations which imply the existence of a large amount of invisible or `dark' matter in the Universe~\cite{RotationCurve,wmap,planck}, but as of yet its composition remains unknown. It is widely expected to be composed of new, massive particles beyond the standard model. Axions are a hypothetical light boson which arise from a solution to the Strong CP problem in quantum chromodynamics~\cite{PQ1977,Wilczek1978}, and can be formulated as a compelling dark matter candidate~\cite{Sikivie1983b,Preskill1983}. Axions are expected to have some mass, and only very weak couplings to the standard model particles. There are various experimental programs around the world which seek to detect axions as dark matter in the local galactic halo. Perhaps the most mature class of detector is the axion haloscope, initially proposed by Sikivie~\cite{Sikivie83haloscope,Sikivie1984}. Axion haloscopes seek to exploit the putative axion-photon coupling, in which an axion interacts with two photons. A typical haloscope experiment consists of an electromagnetic resonant cavity coupled to a low-noise readout chain, embedded in a strong DC magnetic field. If dark matter is composed of axions, and axions have a diphoton coupling as is predicted in many QCD axion models~\cite{K79,Zhitnitsky:1980tq,DFS81,SVZ80,Dine1983}, some of the dark matter axions will interact with the sea of virtual photons provided by the strong magnetic field and convert to real photons with a frequency proportional to their mass according to

\begin{equation}
hf_a=m_a c^2 + \frac{1}{2}m_a v_a^2.
\end{equation}

Here $f_a$ is the real photon frequency, $m_a$ is the axion mass, $v_a$ is the axion velocity, $c$ is the speed of light, and $h$ is Planck's constant. If $f_a$ matches the frequency of a suitable resonant mode inside the resonator, the generated photons are trapped, and may be read out as excess power inside the resonator. Under the typical set of assumptions, the amount of signal power expected due to axion-photon conversion inside a resonator is given by~\cite{Kim2020}

\begin{equation}
P_{a\rightarrow\gamma} = \left(g_{a\gamma\gamma}^2\frac{\rho_a}{m_a}\right)\left(\frac{\beta}{1+\beta}B_0^2VC\text{min}(Q,Q_a)\right).
\end{equation}

Here $g_{a\gamma\gamma}$ is the axion-photon coupling strength, $B$ is the magnetic field strength, $V$ is the volume of the resonant cavity, $\beta$ is the resonator coupling coefficient, $Q$ is the resonant cavity quality factor, $Q_a$ is the effective axion signal quality factor which arises due to the axion velocity distribution broadening the signal in frequency space ($Q_a$ is typically taken to be $\sim10^6$), and $\rho_a$ is the density of axion dark matter. $C$ is a resonant mode-dependant quantity known as the axion haloscope form factor which relates to the overlap of a given resonant mode with the externally applied DC magnetic field. It is defined as~\cite{Sikivie83haloscope}

\begin{equation}
C=\frac{\left(\int \vec{E}\cdot\vec{B}_0~dV\right)^2}{\int \vec{B}_0\cdot\vec{B}_0~dV~\int \vec{E}\cdot\vec{E}~dV}.
\end{equation}

Here $\vec{E}$ is the electric field of the resonant mode in question, and $\vec{B_0}$ is the applied external DC magnetic field. Typically in haloscopes, solenoids with magnetic field lines aligned with the $z$-axis of a cylindrical resonant cavity are employed, and as such only TM (transverse magnetic) modes have non-zero form factors, with the TM$_{010}$ mode in the cylindrical cavity having the highest form factor ($C\approx0.69$ for an empty cavity). Since the axion mass is unknown, and thus the corresponding generated photon frequency is unknown, axion haloscope resonators must be tunable. This is typically done by introducing a cylindrical tuning rod into the resonant cavity, which is then moved radially from the wall to the centre of the cavity, pushing the mode frequency upwards. This technique is very mature in the frequency range around 1 GHz where several haloscopes exist (e.g. ADMX, CAPP) ~\cite{Du_ADMX,Braine2020,Bartram2021,ADMX2021,CAPP_DFSZ_2023,CAPP2021}, but becomes more challenging when moving to higher frequencies such as those of recent theoretical~\cite{SMASH2017,SMASH2019} and experimental interest~\cite{Zhong2018,Brubaker2017b,ORGAN_1a,McAllister201767}. There are several proposals and designs for axion haloscope resonators to enable searches in the higher axion mass regime, and it is a field of ongoing research~\cite{stern_2015,multi_cell,Orpheus_2022,Quiskamp2020,MADMAX2017,Supermode2018}. 

This article considers a relatively simple proposal for a new class of axion haloscope resonator - a rectangular resonant cavity with a tunable wall, which enables tuning the cavity mode frequency without introducing tuning rods, or other elements into the empty resonating chamber. We compare this design with a simple tuning rod cavity, and discuss some advantages of the rectangular design.

\section{Rectangular Resonators}

Similar to hollow cylindrical resonators, rectangular cross-section resonators can support axion sensitive electromagnetic modes. A rectangular conducting cavity as shown in fig.~\ref{fig:rectanglesketch} with dimensions $a$, $b$ and $d$ in the $x$, $y$ and $z$ directions respectively is capable of supporting both TE and TM modes. 

\begin{figure*}[t]
\centering
	\includegraphics[width=0.9\textwidth]{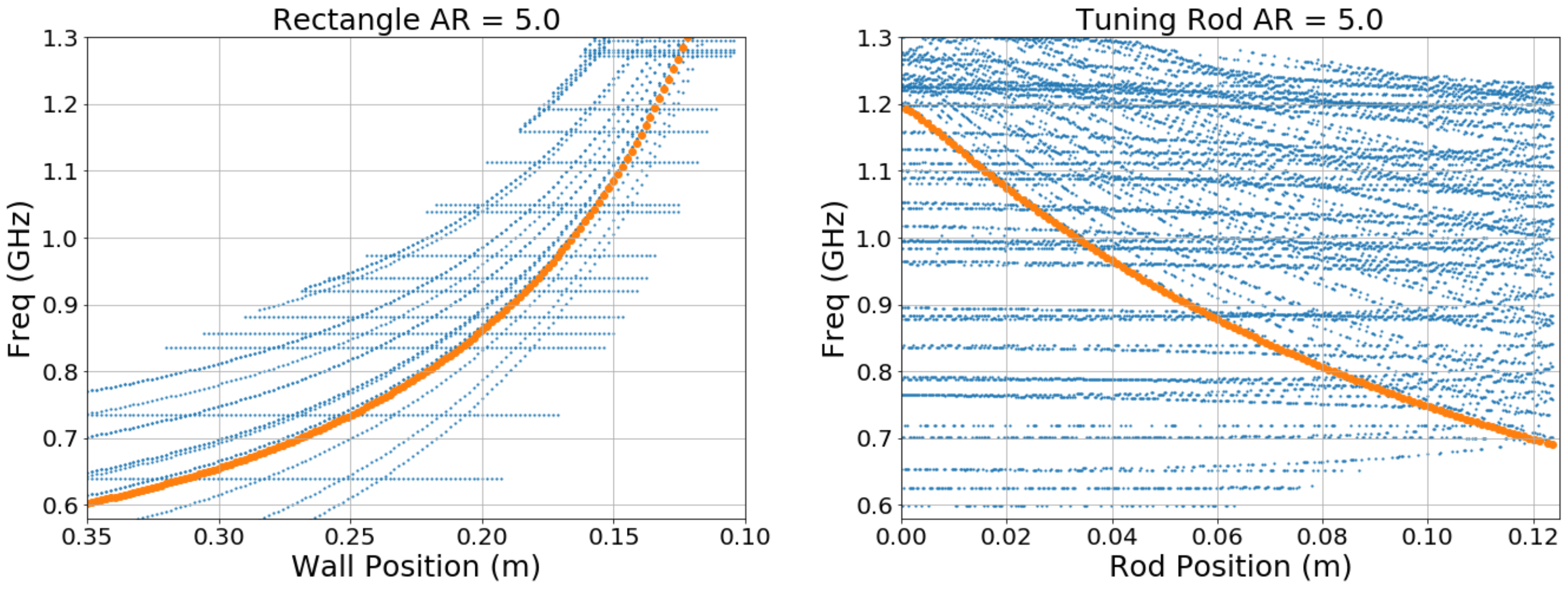}
\caption{Left: Resonant mode frequencies against the tunable wall position for the rectangular haloscope resonator modelled with an aspect ratio of 5, the axion sensitive TM$_{110}$ mode is highlighted in orange, whereas the blue dots represent intruding modes. We note that the frequency tuning range of the rectangular resonator is much greater than what is displayed here, it is truncated for the purposes of a direct comparison to the tuning rod resonator, to display the number of mode crossings present in each case in the tuning range accessible to the tuning rod cavity. Right: Resonant mode frequencies against the tuning rod position for the cylindrical tuning rod haloscope resonator modelled with an aspect ratio of 5, the axion sensitive TM$_{010}$ mode is highlighted in orange, whereas the blue dots represent intruding modes.}
\label{fig:freq}
\end{figure*}

For the purposes of this study, we consider the TM$_{mnp}$ modes with field resonant field patterns given by

\begin{align*}
E_z &= E_0\sin{k_x x}\sin{k_y y}\cos{k_z z},\\
E_x &= \frac{-E_0 k_z k_x}{k^2-k_z^2}\cos{k_x x}\sin{k_y y}\sin{k_z z},\\
E_y &=  \frac{-E_0 k_z k_y}{k^2-k_z^2}\sin{k_x x}\cos{k_y y}\sin{k_z z},\\
H_z &= 0,\\
H_x &= \frac{i E_0 \epsilon_0 k_y}{k^2-k_z^2}\sin{k_x x}\cos{k_y y}\cos{k_z z},\\
H_y &= \frac{-i E_0 \epsilon_0 k_x}{k^2-k_z^2}\cos{k_x x}\sin{k_y y}\cos{k_z z}.
\end{align*}

Here $E$ is the electric field and $H$ is the magnetic field, $E_0$ is a constant which defines the amplitude of the field, $i$ is the imaginary number and $\epsilon_0$ is the vacuum permittivity, $k_x=\frac{m\pi}{a}$, $k_y=\frac{n\pi}{b}$, $k_z=\frac{m\pi}{d}$ and $k^2=k_x^2+k_y^2+k_z^2$. The frequency of a given mode can be found from $2\pi f = kc$. The TE field patterns have similar analytical field expressions, and can be found in textbooks. We note that for TE modes, $m=0$ or $n=0$ is permitted, whereas for TM modes only $p=0$ is permitted. In particular, as shown in fig.~\ref{fig:rectanglesketch} we can see that the TM$_{110}$ mode bears a resemblance to the TM$_{010}$ mode in a cylindrical resonator, in that it has a uniform electric field in the $z$-direction, with no other electric field components. The TM$_{110}$ mode is also non-degenerate (for $a\neq b$). Indeed, if we align the $z$-axis of such a cavity with an externally applied uniform DC magnetic field, we can analytically determine that the form factor for this mode is $C\approx0.66$, close to the empty cylindrical resonator TM$_{010}$ mode value. 

Both of these modes are highly uniform, with the frequency depending only on the radius of the cylinder, or the effective radius of the rectangular resonator (the width and depth, $a$ and $b$). This uniformity is the challenge in tuning such resonant modes, and the reason for introducing tuning rods inside TM$_{010}$ cylindrical haloscopes. It is not typically possible to tune the radius of a cylindrical resonator, on the other hand, it is easier to imagine tuning the width or depth of a rectangular cross section, by having one movable wall mounted to a stepper motor, and moving it inwards, thus reducing for instance the $a$ dimension. 

This kind of design presents a few immediate advantages - the form factor of the rectangular TM$_{110}$ mode is unchanged as the width changes, whereas introducing a tuning rod changes the form factor of a cylindrical resonator. Also, without the need to introduce a tuning rod, we can regain some of the volume lost by moving from a circular cross-section to a rectangular one. 

Tuning the position of a side wall, and thus reducing the effective radius of the rectangular waveguide causes rapid frequency tuning, enabling large tuning ranges within a single cavity with a small range of motion. Additionally, at higher frequencies, as cavities get smaller various technical issues with tuning rods in haloscopes can become exacerbated by the smaller physical sizes. For example it can become difficult to align and rotate small tuning rods inside small, cryogenic resonant cavities (tuning rod jamming already being an issue even at larger cavity sizes~\cite{RodSticking}), whilst having a structure with a single moving plunger or wall can be experimentally easier to implement, and has been done in the field of microwave filters for decades~\cite{plunger1,plunger2,plunger3,plunger4}.

Rectangular cross-sections have been considered before in haloscopes~\cite{RADES,RADESRectangle}, but not in this context, and not with this relatively simple tuning mechanism to our knowledge. For these reasons, we consider the sensitivity of this relatively simple design to axions, and compare it to standard tuning rod-based haloscope resonator designs. 

\section{Haloscope Resonator Design Considerations}

\begin{figure*}
\centering
	\includegraphics[width=0.9\textwidth]{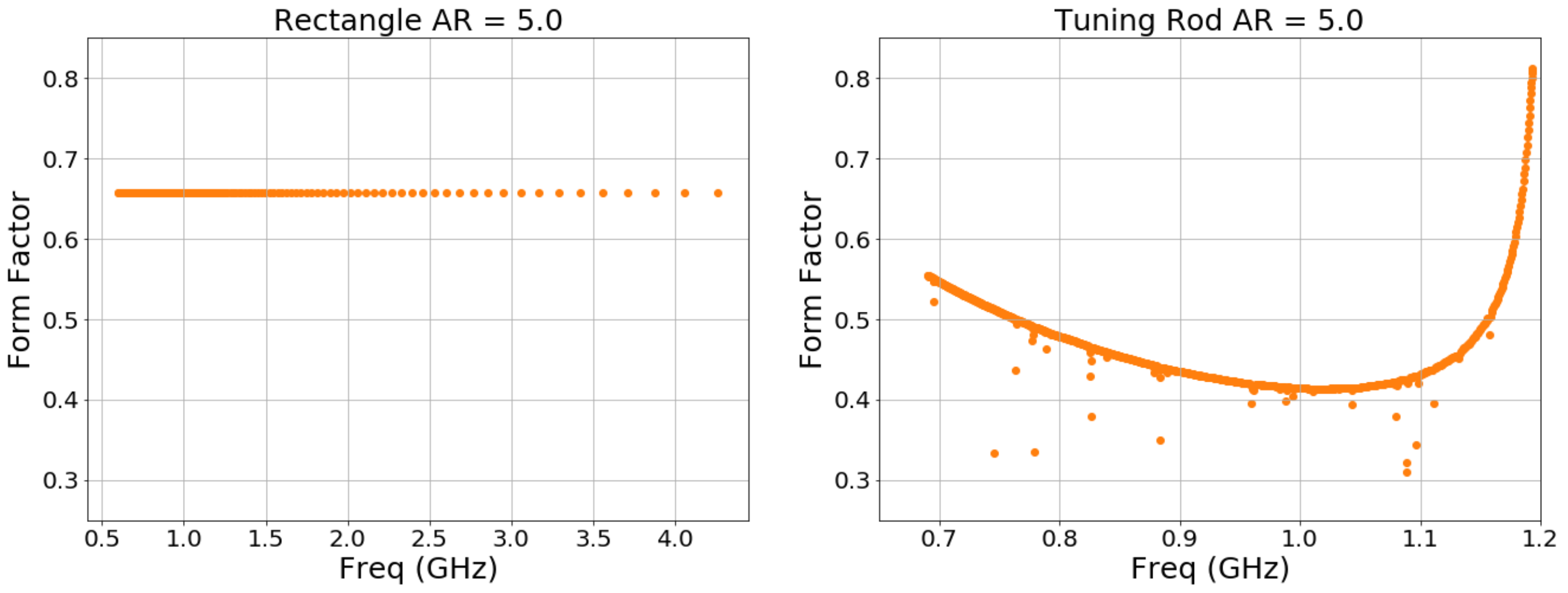}
\caption{Left: Axion haloscope form factor against resonant mode frequency for the rectangular haloscope resonator modelled with an aspect ratio of 5, for the axion sensitive TM$_{110}$ mode. We note that the point density becomes lower at the high frequency end - this is due to the fact that the tuning becomes much more rapid with small wall displacement in this range, and we modelled with fixed wall position steps. Regardless, the rectangular resonator form factor is constant, and the range with low point density is far beyond the tuning range of the tuning rod resonator, where the cavity volume is small and thus outside the likely useful region of the resonator. Right: Axion haloscope form factor against resonant mode frequency for the cylindrical tuning rod haloscope resonator modelled with an aspect ratio of 5, for the axion sensitive TM$_{010}$ mode.}
\label{fig:C}
\end{figure*}

It can be shown that the critical parameter in axion haloscope design is the allowable rate of frequency scanning for a given set of experimental parameters. This quantity is given by~\cite{stern_2015}

\begin{equation}
\frac{df}{dt}\propto\frac{1}{SNR^2}\frac{g_{a\gamma\gamma}^4\rho_a^2Q_a}{m_a^2}\frac{B^4}{(k_bT)^2}C^2V^2Q
\end{equation}

Here $\frac{df}{dt}$ is the scan rate, $SNR$ is the desired signal to noise ratio of the experiment, $k_B$ is the Boltzmann constant, and $T$ is the total system noise temperature. The quantities that can be engineered in a resonant cavity specifically (as opposed to the other components of the experiment such as the magnet, amplification and readout, or the quantities fixed by nature) are the form factor, $C$, the resonant quality factor, $Q$, and the resonator volume, $V$. As such, the figure of merit for haloscope resonator design is $C^2V^2Q$. When computationally modelling resonators, it is common to predict $Q$ factors using a quantity known as geometric factor, $G$, which measures the amount of the electromagnetic field of the resonant mode that is coupled to the conducting walls, inducing currents in those walls and causing loss. G is given by

\begin{equation}
G=\frac{2\pi f \mu_0 \int |\vec{H}|^2~dV}{\int |\vec{H}|^2~dS}.
\end{equation}

Here $\mu_0$ is the vacuum permeability, and $S$ is the cavity surface. $G$ is related to $Q$ by

\begin{equation}
Q=\frac{G}{R_s}.
\end{equation}

Here $R_s$ is the surface resistance of the material that makes up the cavity walls. For the purposes of the coming comparisons, we will consider the figure of merit for haloscope resonant cavity design to be $C^2V^2G$. Another critical design consideration for axion haloscopes is the number of mode crossings, or non-axion sensitive modes that crowd the range of frequency tuning and make experiments more complex. These are typically length dependent modes, which do not tune in the same way as the axion sensitive modes, and thus sometimes coincide in frequency with the axion sensitive mode as the cavity is tuned. To minimise this, tuning rod-based resonators typically employ an `aspect ratio' (the ratio of the cavity length to the cavity radius) of no more than 5. In the coming comparisons we will consider a range of aspect ratios for the rectangular resonator, to determine how they fare in terms of mode crossings. For this purpose we define the effective radius of the rectangular resonator as the distance from the centre to the corner of the cross-section, with the cavity at its largest cross-section before the wall has begun to tune. This can be calculated as

\begin{equation}
r_{\text{effective}}=\frac{\sqrt{a^2+b^2}}{2}.
\end{equation}

The aspect ratio is then defined as

\begin{equation}
AR=\frac{d}{r_{\text{effective}}}.
\end{equation}

Here, again, $a$, $b$ and $d$ refer to the cavity dimensions with the cavity at its largest, with a square cross-section such that $a=b$.

\section{\label{sec:modeling}Finite Element Analysis}

\begin{figure*}[t!]
\centering
	\includegraphics[width=0.9\textwidth]{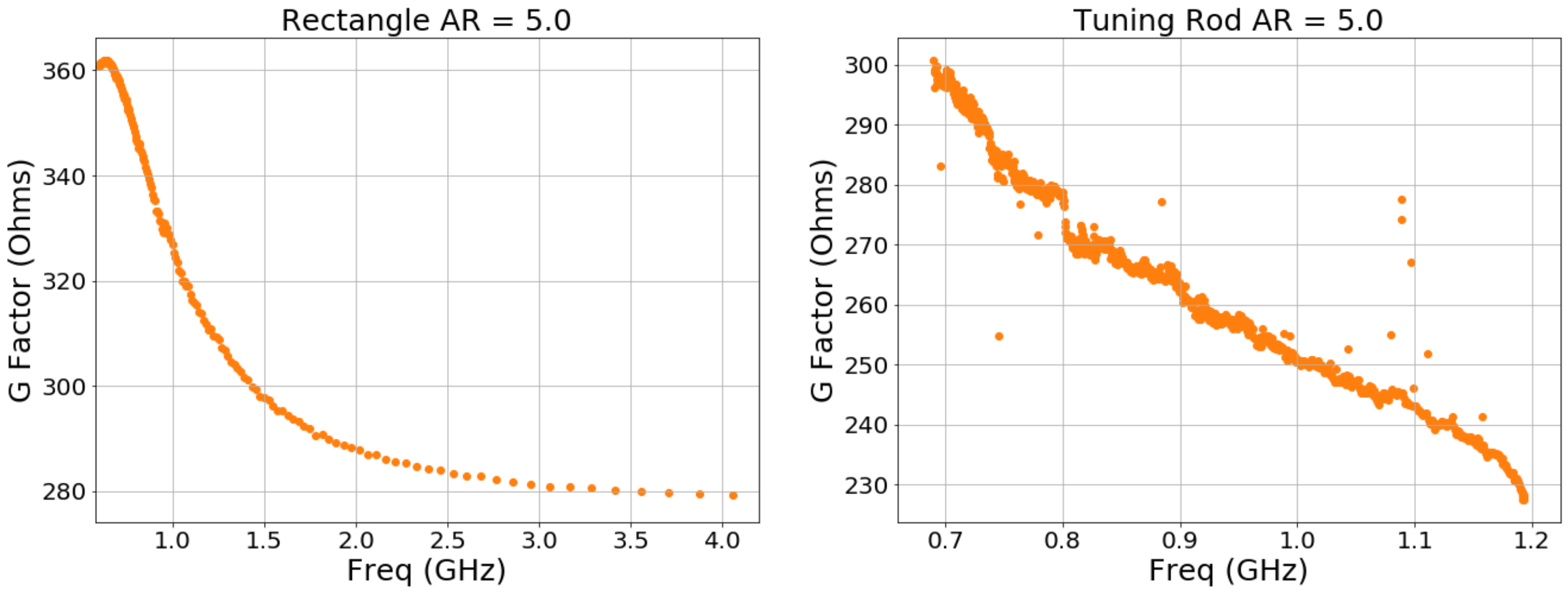}
\caption{Left: Geometric factor against resonant mode frequency for the rectangular haloscope resonator modelled with an aspect ratio of 5, for the axion sensitive TM$_{110}$ mode. We note that the point density becomes lower at the high frequency end for the same reason as in fig.~\ref{fig:C}. Right: Geometric factor against resonant mode frequency for the cylindrical tuning rod haloscope resonator modelled with an aspect ratio of 5, for the axion sensitive TM$_{010}$ mode.}
\label{fig:G}
\end{figure*}

Whilst analytical solutions can easily be found for the frequencies, form factors and geometric factors of rectangular cavity modes and empty cylindrical cavities, they are harder to find for tuning rod-based cylindrical cavities. 

For the purposes of direct comparison, and to analyse the effects of mode interactions, we constructed a finite element model in COMSOL of both types of resonator, to compare the sensitivity of the designs in an arbitrarily chosen frequency range. For an experimentally useful comparison, we consider cavities which would fit inside the ADMX cylindrical magnet bore with a radius of 25 cm, but note that these results apply to any bore diameter, with appropriate re-scaling.

As such, the cylindrical cavity modelled has a radius of 25 cm, and is given an aspect ratio of 5, in keeping with the standard in the field~\cite{stern_2015}. A cylindrical tuning rod is introduced which runs the length of the cavity with no gaps, and is tuned from the wall to the centre of the cavity. The tuning rod radius is half of the cavity radius, as is also common in single rod designs. 

The rectangular cavity is initially given a square cross-section, filling as much of the bore as possible, with a side length of $a=b=50/\sqrt{2}$ cm (or effective radius of 25 cm). The width parameter, $a$ is adjusted from the initial value to one tenth of the initial value, whilst the other dimensions ($b$ and $d$) are left constant. A range of aspect ratios from 2 to 10 are modelled.

We employ the electromagnetic waves package in COMSOL, and find eigenfrequency solutions for the above geometries, using perfect electrical conducting boundaries, and vacuum for the bulk. Radiation losses are not considered for either cavity, nor are the impacts of longitudinal symmetry breaking, which has been shown to increase the coupling of the axion sensitive mode to intruding modes~\cite{Stern2019}. As such, both cavities are modelled in their ideal scenario.

\section{Results, Comparisons and Discussion}

\begin{figure*}[t]
\centering
	\includegraphics[width=0.9\textwidth]{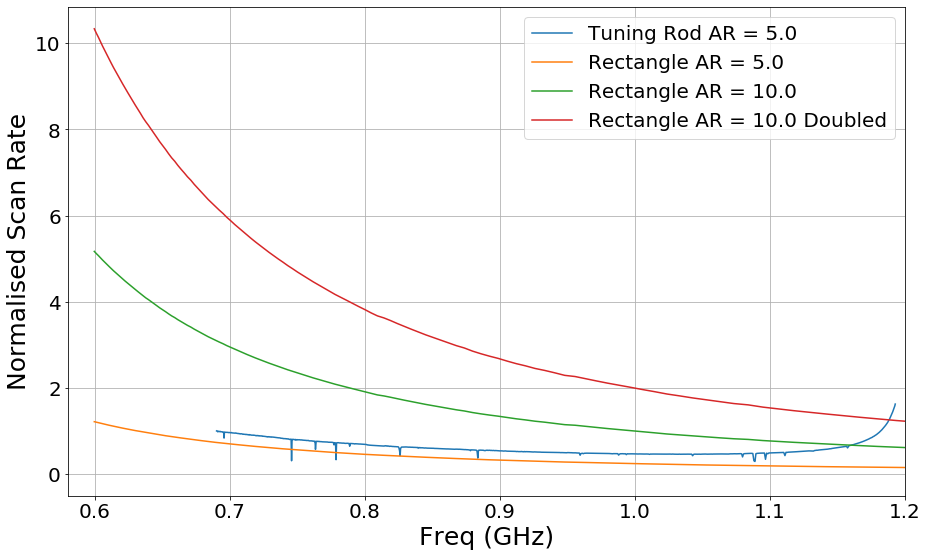}
\caption{Scaled effective scan rates for the various resonators as discussed in the text. The $C^2V^2G$ values in each case have been normalised by the initial $C^2V^2G$ of the tuning rod resonator with the tuning rod at the wall.}
\label{fig:scan}
\end{figure*}

Fig.~\ref{fig:freq} shows frequencies against the relevant tuning parameter for the two designs considered, both with an aspect ratio of 5. As we can see, the rectangular resonator over this range encounters fewer intruding modes for the same aspect ratio. It was not until the aspect ratio was doubled, to a value of 10, that the number of intruding modes in the rectangular cavity became similar to the cylindrical cavity with an aspect ratio of 5. Whilst an aspect ratio of 10 may be infeasible for cavities of the dimension modelled here (as they would be 2.5 m long), at higher frequencies, where the cavity dimensions are smaller, this is promising as it means that rectangular cavities may be constructed with higher aspect ratios than tuning rod cavities, increasing the volume and thus sensitivity to axions without an associated increase in the number of intruding modes. We remind the reader that the frequency band chosen here is essentially arbitrary, selected for direct comparison with ADMX-like experiments, and all parameters can be scaled in frequency as desired by simply scaling the length dimensions of all of the resonators studied.

Fig.~\ref{fig:C} shows axion haloscope form factors against resonant frequency for the two designs considered, again with an aspect ratio of 5. We note that the form factor for the cylindrical resonator experiences several dips over the tuning range, as the TM$_{010}$ mode couples to various intruding modes, hybridising and losing sensitivity to axions as the resonant structure changes. We do not observe this behaviour in the rectangular cavity, indicating a possible lower degree of coupling between the TM$_{110}$ mode and the intruding modes. As stated above, for the purpose of a fair comparison both cavities have been modelled in the `perfect' symmetry case, with no longitudinal symmetry breaking. Of course, in a real resonator there will be some asymmetries, and some mode couplings are expected, but the lack of form factor dips when tuning through intruding modes in this ideal case for the rectangular resonator may indicate lower coupling between the axion sensitive mode and intruding modes than in tuning rod resonators. This could imply narrower avoided level crossings, and smaller forbidden regions of axion sensitivity.

Fig.~\ref{fig:G} shows geometric factors (proportional to resistive $Q$ factors) against resonant frequency for the two designs considered, again with an aspect ratio of 5. We note that the $G$ factors between the two resonators are similar and in fact slightly higher for the rectangular cavity, indicating that the wall losses would be expected to be similar in both cases. Radiation leakage is a matter of consideration for the rectangular cavity, as the presence of a moving wall indicates the requirement for a larger gap in the resonating structure than the axle required for a tuning rod resonator, but this effect can be mitigated to some degree with the use of RF chokes.

Fig.~\ref{fig:scan} shows the effective scan rate figure of merit, $C^2V^2G$ against resonant frequency for both designs considered, under a variety of assumptions. Firstly, the tuning rod is presented with an aspect ratio of 5, along with the rectangular cavity with an aspect ratio of 5. For a more experimentally useful comparison, we also present the rectangular cavity with an aspect ratio of 10, as discussed this cavity has a similar number of intruding modes to the tuning rod cavity with an aspect ratio of 5 over the same frequency range.

\begin{figure*}[t!]
\centering
	\includegraphics[width=0.7\textwidth]{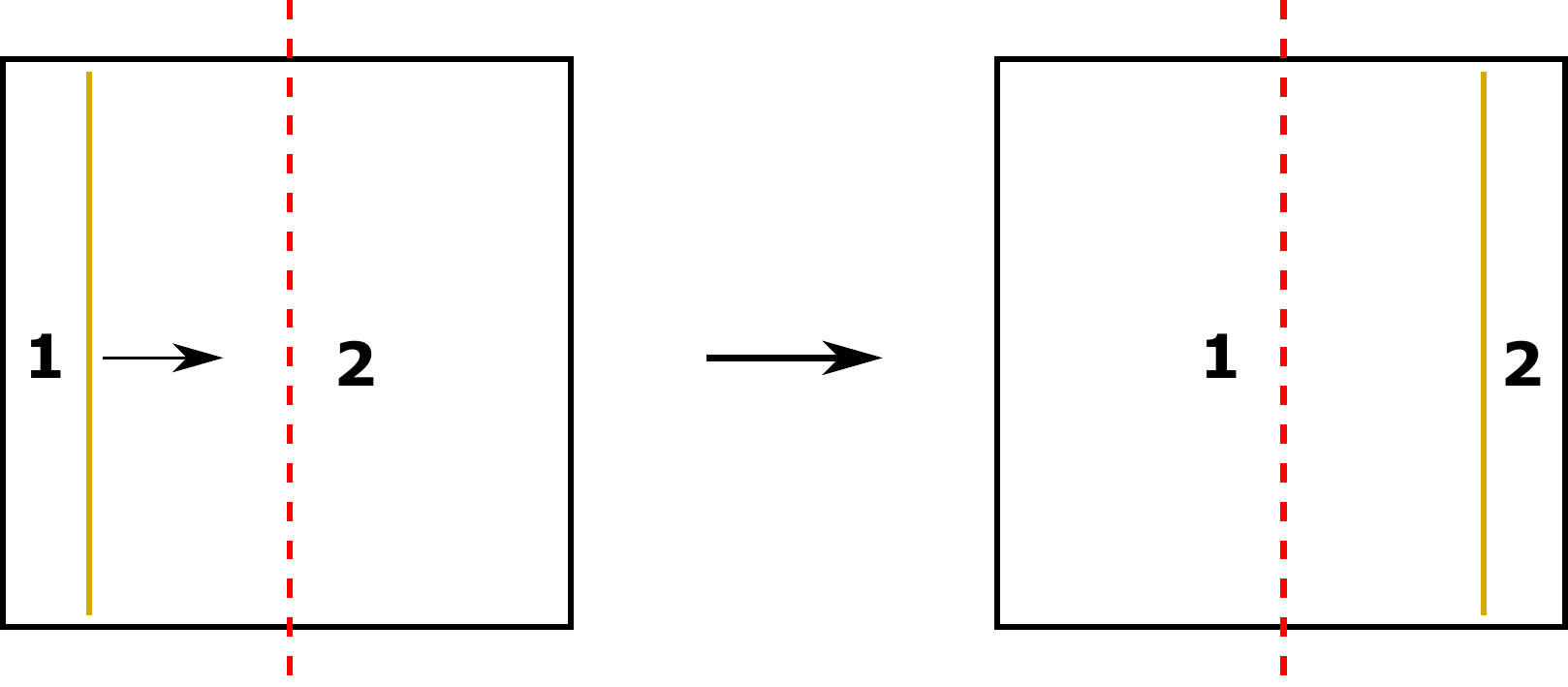}
\caption{A simple schematic of the `doubling' scheme described in the text. On the left we see the initial state of the system in such a scheme, with the tunable wall (gold) close to one side of the outer wall (black). In this case there are two sub-cavities, labelled 1 and 2. As the tunable wall moves past the centre line (dashed, red), and arrives at a symmetrical position close to the other wall of the outer wall, the cavity labelled 1 gets larger as the cavity labelled 2 gets smaller. By the end of the tuning range, the two cavities have occupied an identical set of sizes (and associated frequencies/form factors in the ideal case), albeit at different times. The data from these two cavities may be combined, and can be thought of as effectively doubling the scan rate of the system as each frequency is covered twice, once by each cavity. Gap sizes are exaggerated to make clear the difference between the moving wall and the outer walls.}
\label{fig:2rect}
\end{figure*}

Finally, we present a curve for the `doubled' sensitivity of this higher aspect ratio rectangular resonator. This refers to the effective sensitivity that would be possible in a particular configuration of a rectangular cavity haloscope experiment. As we tune one of the walls of the rectangular resonator, we can effectively create a second cavity without any loss of magnet bore diameter, as shown diagramatically in fig.~\ref{fig:2rect}. If we started the tunable wall in some position on one side of the central point of the available space, coupled readout chains to both rectangular resonators, and tuned the wall through the central point of the available space, stopping in a symmetrical position on the other side, once we had finished we would effectively have had two resonators with the same set of frequencies and form factors, albeit at different times, thus effectively doubling the scan rate of the setup. This is a unique benefit of the rectangular design which comes `for free'. In order to achieve a similar effect in the tuning rod cavity, the cavity would need to be made much smaller so that a second cavity could be added within the same magnet bore diameter. Both designs could benefit equally from axial stacking of cavities, so the effect of this is not considered. We must acknowledge that the `doubling' idea presented here would require a second readout chain and data acquisition system - here we seek only to compare the designs in terms of what is possible with regard to utilising magnet bore cross-sectional space. There would be additional technical challenges associated with such a design - the tunable wall would need to be mounted on a post or rail system which would impact field patterns slightly - so it is best to consider the `doubled' sensitivity as an upper limit of what is possible using this technique if some technical challenges were overcomes, rather than a guaranteed enhancement.

As discussed above, tuning rods are sometimes a technical issue in axion haloscopes as they have a propensity for `sticking' and jamming~\cite{RodSticking}. The gaps between the rods and end caps are a point of significant design consideration. The gaps must be as small as possible whilst still allowing motion to minimise the possibility of longitudinal symmetry breaking (which can cause mode interactions~\cite{Stern2019}), and modes can become localised in these gaps reducing form factor and sensitivity~\cite{ModeLocal1,ModeLocal2}. Tuning rods are also often difficult to thermalise with the resonator and the rest of the system~\cite{HotRod}. These factors can worsen at higher frequencies where necessarily small cavity sizes mean that the machining, assembly and alignment tolerances required to achieve appropriately small relative gap sizes and clearances between moving parts become infeasibly small. These various factors could be mitigated to some extent by removing the tuning rod entirely.

Lastly, when discussing advantages of the rectangular design, we note that the design features a considerable tuning range - we stopped simulating at one tenth of the original cavity width, but in principle this could be extended further. The cavity frequency is extremely sensitive to the tunable wall position, and a single rectangular cavity can access a frequency range many times greater than the range accessible to a tuning rod resonator of similar size - albeit with a reduced sensitivity as the volume decreases.  

This wide range of possible frequencies, and rapid rate of tuning with small mechanical displacement may be interesting to experimentalists who wish to take long data runs without swapping out resonators, or who have limited space for tuning mechanisms. 

It also presents the opportunity to consider designs for specific frequency ranges that look quite dissimilar to the cavities presented here. For instance, cavities with initial cross-sections that are far from square, designed to tune over a targeted frequency band requiring only a very small range of wall positions. Such cavities may have high aspect ratios, specifically selected to minimise mode crossings over the targeted range. The ease of analytical calculations of the mode frequencies in these resonators means this process may be done algorithmically, where software can be used to loop through thousands of possible $a$, $b$ and $d$ values in seconds, and determine the optimal dimensions for a cavity to cover a given frequency range. This approach presents a significant design advantage over the finite element modelling that is currently typically employed in tuning rod resonators. It has already been employed in designing resonators for The ORGAN Experiment.

\section{Experimental Results}
As discussed, axion searches suffer substantial sensitivity and practicality issues with increasing frequency (axion mass), with $\frac{df}{dt} \propto f^{-14/3}$ \cite{stern_2015}.

Rectangular haloscopes offer some relief to these issues, as it is possible to have larger aspect ratios (regaining some of the volume loss at higher frequencies) without an increased number of intruding modes, and a reduced mechanical complexity. To validate this hypothesis, a rectangular cavity has been designed, tested and implemented in The ORGAN Experiment, a high mass axion haloscope hosted at The University of Western Australia.

Phase 1b of The ORGAN Experiment targeted its axion search in the $26 - 27$ GHz ($107.4 -111.9 \,\mu$eV) region of the axion mass-coupling parameter space, becoming the most sensitive high mass axion search to date. The experiment utilised a rectangular cavity haloscope that achieved frequency tuning of the axion sensitive TM$_{110}$ mode in the same way that has been discussed within this paper - by moving one of the side walls in the direction perpendicular to the applied $\vec{B}$ field.

As alluded to, the simple rectangular geometry allowed us to analytically iterate over many different combinations of cavity dimensions to maximise $C^2V^2G$ whilst minimising the number of mode crossings in the desired frequency region. The Phase 1b experiment was designed to have zero intruding modes over the $26 - 27$ GHz target range, which was possible with the following cavity dimensions; $a = 5.6-6$ mm in the tunable $x$ direction, $b = 27.1$ in the $y$ direction and $d = 77.3$ in the $z$ direction, resulting in an aspect ratio of $\sim 5.58$. 

\begin{figure}[t!]
    \centering
    \includegraphics[width=\linewidth]{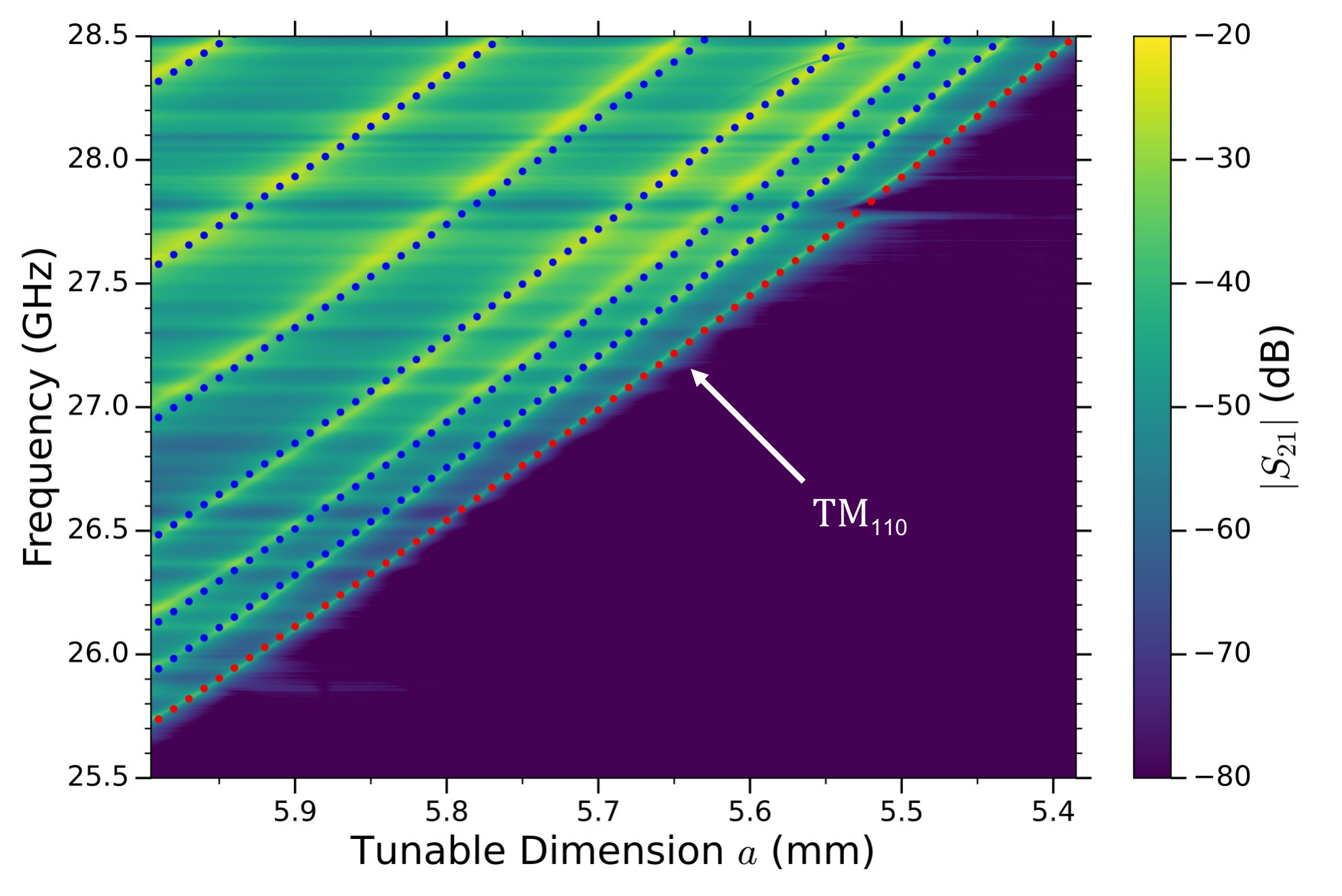}
    \caption{A colour density plot of the transmission coefficient $|S_{21}|$ (dB) as function of frequency and tunable cavity dimension, $a$. The axion sensitive TM$_{110}$ mode is annotated and tunes over the targeted $26-27$ GHz range with no mode crossings. The broad horizontal lines represent standing waves in the cavity-receiver chain. The dots represent the expected positions of the various modes from the COMSOL Multiphysics model of the real cavity.}
    \label{fig:modemap}
\end{figure}

This intruding mode free region is simply not possible with a conventional tuning rod haloscope at the same aspect ratio, demonstrating a benefit of the rectangular haloscope. Frequency tuning of the TM$_{110}$ mode was achieved by moving one of the cavity walls in the $x$ direction via an Attocube ANPz101 piezoelectric stepper motor which was mounted externally to the cavity. The tunable wall had 0.1 mm gaps between all four sides of the wall and cavity and was machined to have a quarter wavelength RF choke to mitigate radiation leakage. Prior to introducing the RF choke, the room temperature $Q$-factor of the cavity was on the order of $\sim1000$ over the tuning range. After introducing the RF choke, the $Q$-factor increased to $\sim5000$ over the range. 

The rectangular haloscope was characterised by taking transmission measurements with a vector network analyzer at different cavity wall positions, which when combined formed a mode map. As shown in fig. ~\ref{fig:modemap}, mode crossings with the target TM$_{110}$ mode and other spurious TE modes are sparse and weakly interacting, except for the avoided level crossing at $\sim 27.8$ GHz. There are no mode interactions in the 26-27 GHz range of interest, and indeed very few in the surrounding region. The broad horizontal lines in the mode map represent standing waves in the cavity-receiver chain and can be attributed to poor impedance matching between the lower frequency, SMA 3.5 mm standard components used and the high frequency cavity mode.

\section{Conclusion}
We introduced the tunable rectangular axion haloscope cavity, and considered it as an alternative to tuning rod resonators in axion haloscopes. We showed that in certain conditions, rectangular haloscopes offer sensitivity benefits over tuning rod designs, and may be of particular use in the push to higher axion frequencies where tuning rod cavities become practically difficult to implement. Additionally, rectangular resonators may be constructed with higher aspect ratios for a given frequency as they contain a smaller number of intruding modes, and associated mode crossings - this is also of particular interest in the high frequency range, where cavities get smaller radially and higher aspect ratios can feasibly be implemented in magnet bores. Finally, we discussed a rectangular axion haloscope that was designed, built and utilised in Phase 1b of The ORGAN Experiment.

\begin{acknowledgments}
This work is supported by Australian Research Council Grants CE17010009 and CE200100008. Ben McAllister is supported by the Forrest Research Foundation. The authors acknowledge Gray Rybka for useful discussions.
\end{acknowledgments}


%

\end{document}